\documentclass[aps,prb,12pt]{revtex4}
\usepackage{graphicx}
\usepackage{amsmath}
\usepackage{amsfonts}
\usepackage{amssymb}
\begin{document}
\title{Einstein-Podolsky-Rosen Paradox in Quantum Diagrams}
\author{S.V.Gantsevich}
\affiliation{Ioffe Institute of Russian Academy of Sciences,
Saint-Petersburg 194021, e-mail: sergei.elur@mail.ioffe.ru}
\begin{abstract}
\baselineskip=2.5ex {\it Quantum diagrams are the best language
for Quantum Mechanics since they show not only a final result but
also the physical process which leads to the result. The quantum
correlation at a distance better known as the
Einstein-Podolsky-Rosen paradox may be easily understood being
depicted in the time-ordered quantum diagrams. In the diagrams one
can clearly see what the so-called entangled quantum states really
are and how they contribute to the violation of Bell inequality.
The wave function phase relations that are the actual physical
"common cause in the past" for the observed correlation become
also visual and evident. Thus the diagram analysis shows that the
phenomenon of distant quantum correlation has simple causal and
local explanation and there is no need to invent various
extravagant constructions contradicting the established physical
principles as well as usual common sense considerations.}
\end{abstract}
\maketitle \baselineskip=2.5ex
Pacs: 71.10.Ca, 72.70.+m, 05.30.-d, 05.40.-a\\
\vskip 0.25 cm \section{Introduction}
\par
In recent decades it was found experimentally that physical
quantities separated by space-like intervals can be correlated.
Logically the correlation between two events in one time moment is
possible either by their mutual interaction or by some "common
cause in the past". However, Relativity Theory forbids the
infinite interaction velocities whereas Quantum Mechanics states
that the values of measured physical quantities do not exist
before the measurements.
\par
Thus it seems that both reasons for the observed correlation
should be excluded if one equally accepts RT and QM. The apparent
contradiction between them was formulated in 1935 by Einstein,
Podolsky and Rosen [\cite{EPR}] being named subsequently the
EPR-paradox. After the immediate Bohr reply [\cite{Br}] defending
the QM consistency the paradox was forgotten for years. In 1951
Bohm proposed its very convenient spin-version [\cite{Bm}] and
then in 1964 Bell constructed his famous inequality [\cite{Bl}]
that made possible the experimental verification of EPR(B) quantum
correlation.
\par
The violation of the Bell inequality in the EPR(B) experiments
[\cite{Asp1,Asp2,Wh,Rw}] proved the validity of QM predictions and
maked actual the adequate interpretation of the quantum
correlation at large distances.
\par
These experiments triggered the enormous incessant flow of
literature with various explanations of the paradox and
interpretations of quantum mechanics (see, e.g.
[\cite{Cav,NPG,WSG,M,R,AFOV,Zr, RBH,Zel}]).
\par
In order to overcome the explanation difficulty one frequently
resorts to various strangely-looking constructions like "momentary
action at a distance", "retrocausality", "multitude of worlds"
etc. All such inventions contradict the traditional physical
picture going from the time of Faraday and Maxwell as well as the
usual common sense.
\par
Recently the very ingenious experiments [\cite{Nat}] confirming
once more with high reliability the violation of Bell inequality
was proclaimed as the final proof of the so-called nonlocality of
QM and the total crash of the point-to-point Faraday-Maxwell
interaction picture that allegedly followed from QM [\cite{Wise}].
\par
Actually the observed correlation is the macroscopic version of
the quantum exchange correlation well-known in the microscopic
quantum systems like atoms, molecules or gases. However, because
of its microscopic character this phenomenon attracted little
attention and the words "it was a quantum effect" usually was
enough for an explanation.
\par
In order to see better the nature and origin of various quantum
effects it is convenient and instructive to represent them in
quantum-mechanical diagrams. The diagrams are the best language
for QM and in some sense their role may be compared with the role
of Leibnitz-Newton Calculus for Classical Mechanics. One may say
without much exaggeration that to study QM without diagrams is the
same as to study Classical Physics not using the differentiation
and integration.
\par
However, the graphics of widely used Feynman diagrams is not fit
for correlation phenomena. The Feynman pictures have no time
ordering of event-points and do not explicitly show the occupancy
numbers of quantum states. All this is highly important for the
understanding of the physics of correlation. Therefore, we shall
use below the kinetic diagrams which we used in the past for
various kinetic problems (see, e.g.([\cite{GGM,FNL}]). What
follows below may be regarded as the graphic illustration of the
analytical formulae of the paper [\cite{GG}] about exchange
correlation phenomena.
\section {Diagrams}
The kinetic diagrams have one-to-one correspondence with the
analytical formulae and can be easily and rather unambiguously
interpreted. We shall not introduce them here in a formal way.
Instead we shall show how to depict them using Feynman diagrams as
prototypes. Since the diagram topology and the meaning of points
and lines are the same in both types of diagrams this can be done
without efforts.
\par
A diagram is the collection of lines with points on them. The
lines correspond to probability amplitudes (or simply to wave
functions). Generally they are the eigenfunctions of the basic
Hamiltonian and are marked by the corresponding indices. The
points relate to the actions of perturbation potentials or to the
observable quantities. The points are proportional to the matrix
elements of corresponding operators. The event-points in kinetic
diagrams have strict time order and in addition to the lines there
are special symbols for the occupancy numbers of quantum states.
\par
Unlike them the Feynman diagrams have no time order for points and
their lines do not distinguish between wave functions and the wave
functions with the occupancy numbers. Because of this a single
Feynman diagram generally describes a number of various physical
processes and in order to treat them adequately some additional
analytical procedures are required. The necessity of additional
analytic calculations as well as the polysemy of Feynman diagrams
hamper their use for the interpretation of correlation phenomena.
\par
To the contrary one kinetic diagrams corresponds as a rule to one
concrete physical process and one final analytical expression.
Hence many kinetic diagrams correspond to one Feynman diagram. The
large number of kinetic diagrams is the price for the simplicity
of rules and the unambiguity of interpretation.
\par
As a simple example let us take the expression for the mean value
of an observable physical quantity, represented by the operator
$A$, in a quantum state $u$ occupied with the probability $F_u$.
\begin{equation}\label{1}
\bar{A}=\langle u|A|u\rangle F_u
\end{equation}
Here $|u\rangle\equiv\psi$ is the wave function of the state $u$
or {\it u-ket} and $\langle u|\equiv\psi^\dag$ is its complex
conjugation or {\it u-bra}. Note that the bra and ket functions
that are necessary to obtain a physical observable are {\it
different quantities with their own time and space dependences}.
The quantity $F_u$ is the occupancy number of the state $u$ or its
distribution function. The values of $F$ lie in the range $[0,1]$
for one observable particle and can be arbitrary for a coarse
grained averaged ensemble of many particles.
\par
Now let us depict this formula as the Feynman and kinetic diagrams
(see Fig.1). The diagrams have only one point so there is no need
for the time ordering (in the picture time goes upwards).
\begin{figure}[htb]
\begin{center}
\includegraphics[width=4in]{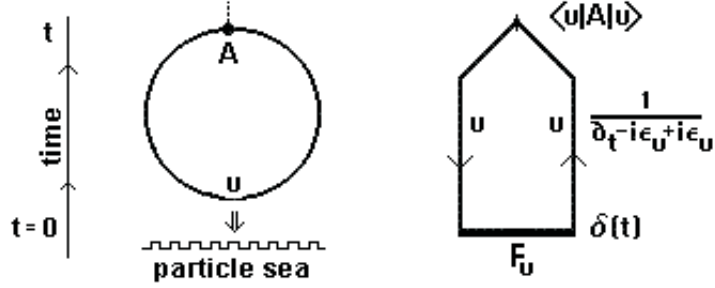}\\
\end{center} \caption{\label{dgepr1} Observed quantity A in quantum state u}
\end{figure}
Looking on the Feynman diagram at the left we see that it does not
fully correspond to the expression (\ref{1}).
\par
At first the occupancy number $F_u$ is absent. Then we see only
one line for the u-bra and u-ket functions in (\ref{1}).
\par
The kinetic diagram at the right has one-to-one correspondence
with the formula (\ref{1}). Reading it from top to bottom (i.e.
against the time) and writing one by one all symbols we have:
\begin{equation}\label{2}
\langle
 u|A|u\rangle
\frac{1}{\partial_t-i\epsilon_u+i\epsilon_u}\delta(t)F_u=\langle
 u|A|u\rangle
\frac{1}{\partial_t}\delta(t)F_u=\langle u|A|u\rangle F_u\Theta(t)
\end{equation}
Here $\Theta(t)=(1/\partial_t)\delta(t)$ is the Heaviside step
function equal to zero at $t<0$ and unity at $t\geq 0$. We see
that in addition to (\ref{1}) the diagram shows the time
dependence of the bra and ket functions.
\par
The resolvent $1/(\partial_t-i\epsilon'+i\epsilon)$ which
corresponds to the bra+ket line section between the observation
point at $t$ and the initial point at $t=0$ is the solution of the
double Schr\"{o}dinger equation governing the combine evolution of
the bra and ket functions.
\begin{equation}\label{3}
\frac{1}{\partial_t-iH'+iH}\overline{\psi^\dag(0)\psi(0)}\delta(t)=
e^{i(H'-H)t}\overline{\psi^\dag(0)\psi(0)}\Theta(t)
\end{equation}
For the eigenfunctions of the Hamiltonian the resolvent contains
only its eigenvalues $\epsilon$ and $\epsilon'$. The energies
(frequencies) of ket-lines enter in the energy denominator with
the plus sign and those of bra-lines with the minus sign. In the
Feynman-type diagram techniques the external frequency $\omega$ or
the Laplace parameter $s$ appear in the resolvent instead of the
time derivative $\partial_t$. The use of $\partial_t$ is
preferable since it combines visual time pictures with the
simplicity of Laplace or Fourier formulae. We can attribute the
symbol $\partial_t$ to the time line and then add it to the bra
and ket energies in the energy denominator. (Note that the
diagrams always represent not time equations but their solutions).
\par
In order to form the occupancy number $F_u$ at $t=0$ the initial
phases of bra and ket should be equal and opposite in signs. The
phase equality survives during time evolution so that the phases
cancel each other and do not appear explicitly in (\ref{1}).
\par
The topology as well as the time or frequency dependence in both
types of diagrams are always the same. Apart from the time
ordering of points the main difference is the presence of the
occupation numbers $F$ in the kinetic diagrams. If in kinetic
diagrams we include these symbols into lines we will get the
corresponding Feynman diagrams.
\par
The transition from a Feynman diagram to the set of corresponding
kinetic diagrams is also simple but requires some practice for
diagrams with many points and lines. After the time ordering of
points the occupation numbers can be introduced by the following
graphic procedure:
\par
Let us imagine that we have a "sea of particles" below $t=0$ in
the diagram (see Fig.1). If we take a line of the diagram and then
plunge it into this sea the line will emerge with the tail-symbol
representing the occupancy number. The tail divides the initial
line into the bra-line and ket-line with the same quantum index
and {\it the same value of intial phase}. In any kinetic diagram
at least one of the lines should have this tail. This simple
graphic procedure may substitute the complicated analytical
correspondence rules which apply to Feynman diagrams when one uses
them for cases where the occupancy numbers are not equal to zero
or unity.
\par
The physical sense of the kinetic diagram in Fig.1 is simple and
transparent. With the probability $F_u$ we have the quantum state
$u$ at $t=0$ formed by the product of u-bra and u-ket. It is
important that their initial (arbitrary) phases coincide and
therefore cancel each other: $e^{+i\varphi_u}e^{-i\varphi_u}=1$.
If u-bra and u-ket have different arbitrary phases their product
will vanish after phase averaging and there will be no permanent
occupation of the quantum state $u$. The occupation number $F_u$
may be regarded as a number of u-bra and u-ket with the coinciding
phases at $t=0$. During time evolution the u-bra and u-ket acquire
(under permanent action of the Hamiltonian) the same phases
$e^{+i\epsilon_ut}e^{-i\epsilon_ut}=1$. (The convolution of these
functions appears as the resolvent in Fig.1). Then the occupancy
number of state $u$ remains unchanged $F_u(t)=F_u(0)$ as well as
the expectation value of the physical quantity $A$ in this state.
\section{Exchange Correlation}
Now let us consider two independent quantum states $u$ and $v$
occupied with the probabilities $F_u$ and $F_v$ and two observable
physical quantities $A$ and $B$. We suppose that these two
quantities are measured by two independent detectors. We suppose
also that we make the series of measurements and can compare their
results made in the same time moment $t$.
\par
Now let us depict the matrix element $\langle uv|AB|uv\rangle
F_uF_v$ which we identify with the jointly averaged product of the
measurement values in both detectors. Two kinetic diagrams for
this quantity are presented in the Fig.2. Two analogous diagrams
can be obtain by the substitution $u\leftrightarrow v$. For
brevity we do not draw them.
\begin{figure}[htb]
\begin{center}
\includegraphics[width=4in]{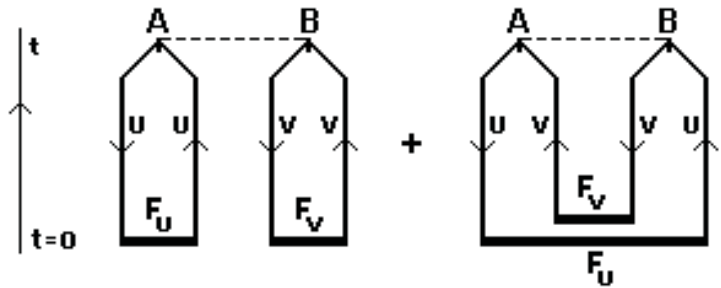}\\
\end{center} \caption{\label{fig:dgepr2} Exchange correlation for two measurements}
\end{figure}
We see in the picture the sum of two diagrams with different
topology. The left diagram consists of two objects with no links
between them (the apical correlation line should not be counted).
The states $u$ and $v$ are represented by the bra and ket lines
with the same indices which form at $t=0$ two independent
occupancy numbers. So we have two pure (bra+ket) pairs. Then one
pair hits one detector while other pair hits another detector.
\par
Reading the diagram we get the product of the independent
contributions of two states:
$$\langle u|A|u\rangle\langle
v|B|v\rangle\frac{1}{\partial_t-i\epsilon_u+i\epsilon_u-i\epsilon_v+i\epsilon_v
}\delta(t)F_uF_v=\langle u|A|u\rangle\langle v|B|v\rangle
F_uF_v\Theta(t)$$ The analogous result $\langle
v|A|v\rangle\langle u|B|u\rangle F_uF_v$ comes from the omitted
diagram with $u\leftrightarrow v$. This way we get the
uncorrelated contribution to $\overline{AB}$ as the sum:
\begin{equation}\label{4}
(\overline{AB})_{uncor}=[\langle u|A|u\rangle \langle v|B|v\rangle
+ \langle v|A|v\rangle \langle u|B|u\rangle ]F_uF_v.
\end{equation}
The diagrams have strict correspondence to the QM rules being
their visualization. We have two measurements and it seems that
these measurements relate also to two objects namely to two
observed quantum particles. Indeed we have two states and two
occupation numbers. But the diagrams clearly show that there are
{\it four entities} represented by {\it four lines}. Each observed
quantum particle is represented by the {\it pair of bra+ket}.
\par
Now let us look at the right diagram of the Fig.2. Its topology
demonstrates quite clearly the correlation between $A$ and $B$
measurements. Here we see also two bra+ket pairs at $t=0$ and two
bra+ket pairs entering the detectors. But unlike the left diagram
now the detector pairs and the occupancy number pairs are
different. The initial $(u,u)$ and $(v,v)$ pairs become $(u,v)$
and $(v,u)$ pairs of the detector devices. The bra and ket have
their own time and space coordinates and just this fact makes the
exchange possible. (In the same way a quantum particle (i.e.
bra+ket pair) can travel through both slits in the two-slit
interference being therefore in one time moment in two different
space points.)
\par
Reading the diagrams from top to bottom and writing consecutively
all symbols we get
\begin{equation}\label{5}
\langle u|A|v\rangle\langle
v|B|u\rangle\frac{1}{\partial_t-i\epsilon_u-i\epsilon_v+i\epsilon_v+i\epsilon_u}
\delta(t)F_u F_v=\langle u|A|v\rangle\langle v|B|u\rangle
F_uF_v\Theta(t)
\end{equation}
The omitted analogous diagram gives the expression with
$u\leftrightarrow v$ so the correlated $\overline{AB}$
contribution is given by
\begin{equation}\label{6}
(\overline{AB})_{cor}=\pm[\langle u|A|v\rangle\langle
v|B|u\rangle+\langle v|A|u\rangle\langle u|B|v\rangle]F_u F_v
\end{equation}
The total average $\overline{AB}$ is the sum of the uncorrelated
and correlated parts:
\begin{equation}\label{7}
\overline{AB}=[\langle u|A|u\rangle \langle v|B|v\rangle \pm
\langle u|A|v\rangle \langle v|B|u\rangle ]F_u
F_v+(u\rightleftarrows v).
\end{equation}
The sign of correlation contributions is positive for bosons and
negative for fermions. For fermions the correlation contribution
diminishes the uncorrelated value e.g. $\overline{AB}\equiv 0$ in
the limit $u=v$. For bosons the situation is inverse and the
correlation increases the uncorrelated value.
\par
The right diagram in the Fig.2 without F-symbols becomes the
well-known Feynman loop-diagram. The loop can be redrawn as the
kinetic correlation diagram by the line plunging procedure
described above in the Fig.1. Also the right connected diagram of
the Fig.2 can be obtain from the left unconnected diagram by the
line exchange. Graphically the exchange of lines inevitably leads
to the their intersection or, one may say, to their {\it
"Verschr\"{a}nkung"} alias {\it "entanglement"}. (For better
transparency the correlation diagram in Fig.2 is depicted in a
more compact form without line intersection.)
\par
The diagrams and the corresponding formulae demonstrate the
existence of the correlation between non-interacting quantum
particles which reveals itself as the correlation of observable
physical quantities. One should emphasize that the correlation
emerges only after the joint averaging of detector data. The data
averaged separately in both detectors give simply the sum of
independent contributions of both states:
\begin{eqnarray}\label{8}
\overline{A}=\langle u|A|u\rangle F_u + \langle v|A|v\rangle
F_v\\\nonumber \overline{B}=\langle u|B|u\rangle F_u + \langle
v|B|v\rangle F_v
\end{eqnarray}
These expressions have no trace of correlation and correspond to
the uncorrelated part $(\overline{AB})_{uncor}$ in (\ref{4}).
\par
Now let us discuss the physical reason for the absence of the
exchange contribution in (\ref{8}) and its presence in (\ref{7}).
\par
These different results follow from the same series of concrete
measurements in both detectors and differ only by the averaging
procedures. Therefore the exchange contribution should be {\it
always present} in the concrete measurement data in both
detectors. It vanishes after separate averaging and emerge after
joint averaging of detector data.
\par
We can see how and why it occurs in the diagrams for the separate
measurements in two detectors (see Fig.3). The sum of diagrams at
the left describes the contributions of $(u,u)$ and $(u,v)$
bra-ket pairs entering the detector $A$ while the diagram sum at
the right corresponds to the analogous contributions of $(v,u)$
and $(v,v)$ pairs into the detector $B$. (The diagrams with
$u\leftrightarrow v$ are analogous and for brevity we do not
depict them).
\par
The left and the right parts of the diagram sums (together with
analogous $u\leftrightarrow v$ diagrams) give the formulae
(\ref{8}). These expressions do not depend on phases and time and
give always the same result either for separate or for joint
averaging of concrete detector data.
\par
The diagram in the middle of the Fig.3 is the dissected
correlation diagram of the Fig.2. It describes the contributions
of the mixed $(u,v)$ and $(v,u)$ bra-ket pairs. The disrupted
F-symbols show that their bra and ket import into detectors
separately their initial amplitudes and phases. They belong to
different occupied states and therefore their separate
contributions to the detectors are time and phase dependent.
\begin{figure}[htb]
\begin{center}
\includegraphics[width=4in]{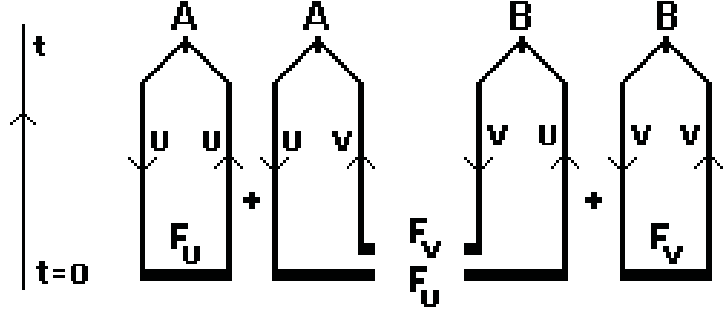}\\
\end{center} \caption{\label{fig:dgepr3} Measurements in separate detectors}
\end{figure}
\par
Reading the middle diagram related to the detector $A$ we get:
\begin{equation}\label{9}
\langle u|A|v\rangle
\frac{1}{\partial_t-i\epsilon_u+i\epsilon_v}\delta(t)\sqrt{F_u
F_v}e^{i\varphi_u}e^{-i\varphi_v}=\langle u|A|v\rangle
e^{i(\epsilon_u-\epsilon_v)t+i(\varphi_u-\varphi_v)}\sqrt{F_uF_v}\Theta(t)
\end{equation}
Here $\sqrt{F_u}$ and $\sqrt{F_v}$ are the amplitudes of bra and
ket while $\varphi_u$ and $\varphi_v$ are their initial phases.
\par
The other part of this diagram gives the analogous expression for
the detector $B$:
\begin{equation}\label{10}
\langle v|B|u\rangle
\frac{1}{\partial_t-i\epsilon_v+i\epsilon_u}\delta(t)\sqrt{F_u
F_v}e^{i\varphi_v}e^{-i\varphi_u}=\langle u|B|v\rangle
e^{i(\epsilon_v-\epsilon_u)t+i(\varphi_v-\varphi_u)}\sqrt{F_uF_v}\Theta(t)
\end{equation}
The expressions (\ref{9}) and (\ref{10}) contain the oscillating
phase multipliers $e^{i\phi(t)}$ and $e^{-i\phi(t)}$ which become
zero after averaging over the initial phases or over time (for
$\epsilon_u\neq\epsilon_v)$. Thus, despite the presence of the
mixed bra-ket pairs in the detectors their contributions do not
appear in the detector data averaged separately. The expressions
(\ref{8}) describe just the result of such averaging.
\par
Quite contrary in the expression (\ref{7}) for $\overline{AB}$
(depicted in Fig.2) the $(u,v)$ and $(v,u)$ mixed pairs
contributions appear as the product with
$e^{i\phi(t)}e^{-i\phi(t)}=1$ and survive under joint averaging of
the detector data.
\par
Thus just this exchange contribution which cannot be observed in
separate detectors is the physical cause of correlation of two
detectors. Let us emphasize that this {\it common cause in the
past} is the real reason for the correlation and not the mythical
"momentary spooky action at a distance" with the mysterious
"mutual influence of measurements". The correlation takes place
for any physical quantities $A$ and $B$.
\par
Two mixed bra+ket pairs of the dissected diagram in Fig.3 are just
the mysterious "entangled quantum particles". They have the same
(and sign-opposite) values of initial phase difference and
conserve them during the time evolution. The mutual phase
coherence of such mixed (bra+ket) pairs can hold at very large
time and space intervals in the absence of perturbations. Just the
presence of such mutually correlated mixed (bra+ket) pairs ensures
the quantum exchange correlation observed at large macroscopic
distances.

\section{EPR(B) spin correlation}
\par
For the EPR(B) spin correlation the operators $A$ and $B$ are the
scalar products of the spin vector $\sigma$ formed by the Pauli
matrices and the arbitrary unit vectors $a$ and $b$. The ket
functions $|u\rangle\equiv|\uparrow\rangle$ and
$|v\rangle\equiv|\downarrow\rangle$ now are the eigenfunctions of
$\sigma_z$ with the eigenvalues $s_\uparrow=1$ and
$s_\downarrow=-1$. In polar coordinates the spin operator $S$ is
given by
$$S\equiv({\bf n,{\sigma}})=\sin\theta
\cos\varphi\sigma_x+\sin\theta\sin\varphi\sigma_y+\cos\theta\sigma_z$$
\par
and its diagonal matrix elements are
\begin{equation}\label{11}
\langle\uparrow|S|\uparrow\rangle=
\cos\theta\langle\uparrow|\sigma_z|\uparrow\rangle=\cos\theta,\quad
\langle\downarrow|S|\downarrow\rangle=
\cos\theta\langle\downarrow|\sigma_z|\downarrow\rangle=-\cos\theta.
\end{equation}
while the nondiagonal elements is given by
\begin{eqnarray}\label{12}
\langle\uparrow|S|\downarrow\rangle=
\cos\varphi\sin\theta\langle\uparrow|\sigma_x|\downarrow\rangle+
\sin\varphi\sin\theta\langle\uparrow|\sigma_y|\downarrow\rangle=
e^{i\varphi}\sin\theta,\\\nonumber
\langle\downarrow|S|\uparrow\rangle=
\cos\varphi\sin\theta\langle\downarrow|\sigma_x|\uparrow\rangle+
\sin\varphi\sin\theta\langle\downarrow|\sigma_y|\uparrow\rangle
=e^{-i\varphi}\sin\theta.
\end{eqnarray}
Using (\ref{11}) and (\ref{12}) we get from (\ref{7}) the
correlation of spins measured in the directions ${\bf n=a}$ and
${\bf n=b}$:
\begin{equation}\label{13}
\overline{AB}=-\cos\gamma 2F_\uparrow F_\downarrow
\end{equation}
Here $\gamma$ is the angle between the measured spin directions
$$\cos\gamma=\cos\theta_a\cos\theta_b+\cos(\varphi_a-\varphi_b)\sin\theta_a\sin\theta_b$$
The first item here comes from the diagram without correlation in
Fig.2 (i.e. from the expressions (\ref{4}) while the second item
comes from the exchange diagram (i.e. from the expressions
(\ref{6}). The sum of them (\ref{7}) becomes independent of the
direction of the spin vector $\sigma$ being proportional to the
scalar product of the unit vectors ${\bf a}$ and ${\bf b}$.
\par
Usually one gets the expression for $\overline{AB}$ as a matrix
element $\langle\Psi|AB|\Psi\rangle$ where $\Psi$ is the
two-particle singlet wave function
\begin{equation}\label{14}
\Psi=(|\uparrow_a\rangle|\downarrow_b\rangle-|\uparrow_b\rangle|\downarrow_a\rangle)/\sqrt{2}
\end{equation}
Using $\Psi$ we get
\begin{equation}
\overline{AB}=\langle\Psi|AB|\Psi\rangle=-\cos\gamma
\end{equation}
and
\begin{equation}\label{16}
\overline{A}=\langle\Psi|A|\Psi\rangle=0,\quad
\overline{B}=\langle\Psi|B|\Psi\rangle=0
\end{equation}
These expressions are equal to the previous expressions under
conditions $F_\uparrow=F_\downarrow$ and $F_\uparrow
F_\downarrow=1/2$.
\par
One frequently treats the identity $\langle\Psi|S|\Psi\rangle=0$
for any spin direction as some peculiar property of the singlet
function (\ref{14}) which is "maximally entangled". The
expressions (\ref{16}) correspond to the expressions (\ref{8}) for
the mean detector data from two independent sources. These
expressions for two spins take the form:
\begin{eqnarray}\label{17}
\overline{A}=\langle\uparrow|A|\uparrow\rangle
F_\uparrow+\langle\downarrow|A|\downarrow\rangle F_\downarrow
=\cos\theta_a (F_\uparrow-F_\downarrow),\\\nonumber
\overline{B}=\langle\uparrow|B|\uparrow\rangle
F_\uparrow+\langle\downarrow|B|\downarrow\rangle F_\downarrow
=\cos\theta_b (F_\uparrow-F_\downarrow).\\\nonumber
\end{eqnarray}
For equal number of opposite spins $F_\uparrow=F_\downarrow$ we
have $\overline{A}=\overline{B}=0$. \par One should also remember
that the correlation is of statistical nature and to observe it a
series of measurements is required. The Malus law for a flow of
polarized spins measured at some angle is realized as random
series of positive and negative units that appear with the
probabilities of positive and negative results:
\begin{equation}\label{18}
\cos\theta=(+1)\cos^2\theta/2+(-1)\sin^2\theta/2=\cos^2\theta/2-\sin^2\theta/2.
\end{equation}
As we see from (\ref{17}) for two spin flows of the opposite signs
the mean observed value is proportional to the difference of
intensities of the flows:
\begin{equation}\label{19}
\overline{S}=\langle\uparrow|S|\uparrow\rangle
F_\uparrow+\langle\downarrow|S| \downarrow\rangle
F_\downarrow=\cos\theta(F_\uparrow-F_\downarrow).
\end{equation}
The observation of spins in the singlet states gives zero mean
value for any measurement direction just because of the parity of
intensities $F_\uparrow=F_\downarrow$. This parity is the unique
real property of the singlet state. For $\theta=\pi/2$ the
detector always gives mean zero irrespective of the flow
intensity. In this case there will be equal average number of
positive and negative units fixed by the detector.
\par
Now let us see how the Bell inequality is violated by the quantum
exchange correlation. We choose two directions ${\bf a}$ and ${\bf
a'}$ of the detector $A$ and two directions ${\bf b}$ and ${\bf
b'}$ of the detector $B$. Suppose that all vectors lay in the
plane $(z,x)$. The vectors ${\bf a}$ and ${\bf b}$ are parallel
and $\theta_a=\theta_b=0$. The vectors ${\bf a'}$ and ${\bf b'}$
are mutually orthogonal and $\theta_{a'}=-\theta_{b'}=\pi/4$. The
Bell sum of $AB$ spin data correlation where the spin values $A$
and $B$ are always equal to $\pm1$ is given by
$$|A(B+B')+A'(B-B')|=|AB+AB'+A'B-A'B'|=|1+1+1-1|=2$$
Taking $AB$-products as $\cos\gamma$ we have for chosen detector
directions the corresponding QM expression:
$$|\cos\gamma_{ab}+\cos\gamma_{ab'}+\cos\gamma_{a'b}-\cos\gamma_{a'b'}|
=|1+1/\sqrt{2}+1/\sqrt{2}-0|=1+\sqrt{2}>2$$ Note that in this
expression the values of first three terms come only from the
uncorrelated part of $\cos\gamma$ in (\ref{13}) because of
$\sin\theta_a=\sin\theta_b=0$ and the correlated part is absent.
The value of fourth term
$A'B'\rightarrow\cos\gamma_{a'b'}=\cos(\pi/2)=0$ is the sum of the
both parts of (\ref{13}) which cancel each other. If we exclude
the correlated contribution the sum will satisfy the Bell
inequality: $|1+1/\sqrt{2}+1/\sqrt{2}-1/2|=1/2+\sqrt{2}<2$. We see
that the Bell inequality is violated just because of the presence
of the mutually correlated contributions in both detectors.
\section{Coulomb exchange interaction}
In the EPR(B) correlation experiments the enormous efforts were
made to exclude any real or imaginal link between detectors. At
last all rivals of the instantaneous (spooky) action at a distance
were removed [\cite{Nat}] and the incontestable nonlocality of QM
(and all nature) was solemnly proclaimed [\cite{Wise}].
\par
Since the observed correlation has no relation to this mythical
phenomenon and can be simply explained by the common cause in the
past, it seems that the "struggle with loopholes" is a little
redundant. The real problem in correlation experiments is to
compare just those detector data that relate to mutually
correlated fluctuations over the permanent uncorrelated
background. For this purpose one can use some real fast
interaction between the mutually correlated events.
\par
Let us consider one such case where two observation points are
connected by real (quasi)momentary interaction. This case is the
Coulomb interaction of charged particles in a quantum system. The
fast Coulomb interaction may connect randomly appearing charges
and serve as the indicator of correlation. The effect can be
depicted in diagrams.
\par
Above we treated the horizontal apical lines in Fig.2 as the
comparison signals for the detector data fixed in the same time
$t$ in both detectors.
\par
Now let these lines represent the Coulomb potential $1/(|{\bf
r}-{\bf r'}|)$ between charges at the points ${\bf r}$ and ${\bf
r'}$. The appearing charges play a role of detecting devices in
these points for the electron states $u$ and $v$. For better
agreement with usual formulae we take the case $F_u=F_v=1$.
Reading the diagrams we get the well-known expression for the
Coulomb interaction energy of two electrons ($e$ is the electron
charge):
\begin{equation}\label{20}
W=e^2\int_V d{\bf r}d{\bf r'}[\frac{|\psi_u({\bf
r})|^2|\psi_v({\bf r'})|^2}{|{\bf r}-{\bf
r'}|}-\frac{\psi_u^\dag({\bf r})\psi_v({\bf r}) \psi_v^\dag({\bf
r'})\psi_u({\bf r'})}{|{\bf r}-{\bf r'}|}]
\end{equation}
The first term corresponds to the left (uncorrelated) diagram in
Fig.2 whereas the second term corresponds to the right
(correlated) diagram. (The contributions of the omitted diagrams
with $(u\leftrightarrow v)$ in Fig.2 are included automatically in
(\ref{20}) because of the double integration over space).
\par
The uncorrelated first term in (\ref{20}) has the simple and clear
meaning. Two charge clouds with the densities $e|\psi_u({\bf
r})|^2$ and $e|\psi_v({\bf r'})|^2$ interact by the Coulomb
potential. The clouds are formed by pure bra+ket pairs and do not
depend on their phases and time. This contribution to the energy
does not differ from analogous classical expressions.
\par
The second term in (\ref{20}) is usually called the exchange
energy and according to QM manuals and books its existence is a
quantum effect which has no classical analogy. Let us note by the
way that the Planck constant $\hbar$ being explicitly written does
not appear in this expression. Also the Coulomb potential
definitely points out that it is the interaction of electric
charges. But what charges? Are they different from the charges in
the first term? And why?
\par
Let us substitute one integration variable in the Coulomb
potential in (\ref{20}) by an external variable ${\bf R}$. The
substitution kills the exchange energy term since $\langle
u|v\rangle=\langle v|u\rangle=0$ (also $\langle u|u\rangle=\langle
v|v\rangle=1$). Then we come to the expression for the energy of
electrostatic field measured by the test charge $e$ at the point
${\bf R}$:
\begin{equation}\label{21}
W({\bf R})=e^2\int_V d{\bf r}\frac{|\psi_u({\bf r})|^2}{|{\bf
r}-{\bf R}|}\quad\mbox{or}\quad W({\bf R})=e^2\int_V d{\bf
r'}\frac{|\psi_v({\bf r'})|^2}{|{\bf R}-{\bf r'}|}
\end{equation}
Here one electron is used as a test charge. An external unit test
charge would show the electrostatic fields $\Phi({\it R})$ created
by two independent charge distributions:
\begin{equation}\label{22}
\Phi({\it R})=e\int_V d{\bf r}\frac{|\psi_u({\bf r})|^2}{|{\bf
R}-{\bf r}|}+e\int_V d{\bf r'}\frac{|\psi_v({\bf r'})|^2}{|{\bf
R}-{\bf r'}|}
\end{equation}
The interaction of two charge distributions (\ref{22}) leads to
the first part of the expression (\ref{20}) which corresponds to
the uncorrelated diagram in Fig.2. The diagram shows that the
charges at points ${\bf r'}$ and ${\bf r}$ are produced by the
pure bra+ket pairs and therefore do not depend on their phases.
\par
Quite contrary in the exchange energy expression in (\ref{20}) the
mixed $(uv)$ and $(vu)$ bra+ket pairs appear together at the
points ${\bf r'}$ and ${\bf r}$. A mixed pair has random phase and
therefore the charge it produces at a given point is the charge
fluctuation with mean zero values. Averaged separately over the
initial phase difference or observation time these charge
fluctuations vanish in each point. But the $(uv)$ and $(vu)$
charge fluctuations are mutually correlated so that while
appearing they have time to interact and give non-zero
contributions to the mean interaction energy. The equal phase
values of these mixed bra+ket pairs cancel each other and do not
enter explicitly in the final expression (\ref{20}). The presence
of correlated random charge fluctuations may be illustrated by the
diagrams in Fig.3.
\section{Conclusion}
The electron exchange correlation as well as many other
microscopic quantum correlation effects have the same physical
nature as the macroscopic EPR-correlation. The resemblance between
this correlation and various microscopic correlation phenomena was
marked from time to time in the literature but usually as the
universal manifestation of nonlocality and instantaneous links
between quantum states.
\par
On the one hand, it is highly doubtful that the application of
such non-physical phantasies can improve the understanding of
microscopic correlation mechanisms. On the other hand, the
abstruse EPR-paradox in reality has simple physical explanation
and just this explanation can help to clarify much more
utilitarian effects.


\begin{references}
\bibitem{EPR} A.Einstein, B.Podolsky, N.Rosen, Phys. Rev., {\bf 47}, 777 (1935).
\bibitem{Br} N.Bohr, Phys. Rev., {\bf 48}, 696 (1935).
\bibitem{Bm} D.Bohm and Y.Aharonov, Phys. Rev. {\bf 108}, 1070 (1957).
\bibitem{Bl} J.S.Bell, Physics 1, 195, (1964)
\bibitem{Asp1} A.Aspect, J.Dalibard, G.Rodger, \prl, {\bf 49}, 91 (1982).
\bibitem{Asp2} A. Aspect, J.Dalibard, G.Rodger, \prl, {\bf 49}, 1804 (1982).
\bibitem{Wh}  G.Weihs et al., Phys. Rev. Lett. 81, 5039 (1998)
\bibitem{Rw} N.A.Rowe et al., Nature 409, 791, (2001)
\bibitem{Cav} N.Brunner, D.Cavalcanti, S.Pironio, V.Scarani, S.Wehner, \rmp, {\bf 86}, 419 (2014).
\bibitem{NPG} M.Navascues and D.Peres-Garcia, \prl, {\bf 109}, 160405 (2012).
\bibitem{WSG} S.P.Walborn, A.Salles, R.M.Gomes et al., \prl, {\bf 106}, 130402 (2011).
\bibitem{M} T.Maudlin, Am.J.Phys, {\bf 78}, 121 (2010).
\bibitem{R} M.D.Reid, P.D.Drummond, W.P.Bowen et al., \rmp, {\bf 81}, 1727 (2009).
\bibitem{AFOV} L.Amico, R.Fazio, A.Osterloh, V.Vedral, \rmp, {\bf 80}, 518 (2008).
\bibitem{RBH} J.M.Raimond, M.Brune, S.Haroche, \rmp, {\bf 73}, 565 (2001).
\bibitem{Zr} W.H.Zurek, \rmp, {\bf 75}, 715 (2003).
\bibitem{Zel} A.Zeilinger, \rmp, {\bf 71}, S288 (1999).
\bibitem{Nat} B.Hensen et al., Nature, {\bf 526}, 682 (2015).
\bibitem{Wise} H.Wiseman, Nature, {\bf 526}, 649 (2015).
\bibitem{GGM} S.V.Gantsevich, V.L.Gurevich, M.I.Muradov et al., Phys.Rev.B, 52, 14006, (1995).
\bibitem{FNL} R.Katilius et al., Fluct.Noise Lett.,{\bf 9},N4, 373, (2010);{\bf 12},N4, 1350023 (2013)
\bibitem{GG} S.V.Gantsevich, V.L.Gurevich, arXiv:quant-ph/1512.03762 (2016).
\end{references}
\end{document}